# Streamlining Knowledge Graph Construction with a façade: The SPARQL Anything project

**Luigi Asprino**
University of Bologna, Italy
luigi.asprino@unibo.it

**Enrico Daga**
The Open University, United Kingdom
enrico.daga@open.ac.uk

**Justin Dowdy**
Independent Software Engineer
justin2004@hotmail.com

**Paul Mulholland**
The Open University, United Kingdom
paul.mulholland@open.ac.uk

**Aldo Gangemi**
Consiglio Nazionale delle Ricerche, Italy
aldo.gangemi@cnr.it

**Marco Ratta**
The Open University, United Kingdom
marco.ratta@open.ac.uk

**Abstract**

What should a data integration framework for knowledge engineers look like? Recent research on Knowledge Graph construction proposes the design of a *façade*, a notion borrowed from object-oriented software engineering. This idea is applied to SPARQL Anything, a system that allows querying heterogeneous resources *as-if* they were in RDF, in plain SPARQL 1.1, by overloading the SERVICE clause. SPARQL Anything supports a wide variety of file formats, from popular ones (CSV, JSON, XML, Spreadsheets) to others that are not supported by alternative solutions (Markdown, YAML, DOCx, Bibtex). Features include querying Web APIs with high flexibility, parametrised queries, and chaining multiple transformations into complex pipelines. In this paper, we describe the design rationale and software architecture of the SPARQL Anything system. We provide references to an extensive set of reusable, real-world scenarios from various application domains. We report on the value-to-users of the founding assumptions of its design, compared to alternative solutions through a community survey and a field report from the industry.

## 1 Introduction

What should a data integration framework for knowledge engineers look like? Approaches can transform the non-RDF data sources on the basis of specific ontologies, designed to represent content from popular formats (e.g. Any23). Alternatively, solutions could offer a mapping language (e.g. RML) which reuses components of format-specific query languages (e.g. JSONPath). In addition, systems could extend SPARQL, and incorporate features of pre-existing query languages for each one of the original formats (e.g. Xpath for XML), allowing users to perform mappings within SPARQL queries (SPARQL Generate). In this resource paper, we present a system whose design principle comes from the notion of façade, borrowed from object-oriented software engineering. A façade is a generic interface that aims at hiding the internal complexity of a class, exposing behaviours that better fit the task at hand. This idea, originally introduced by (Daga et al. 2021), is applied to SPARQL Anything, a system that allows querying heterogeneous resources as-if they were in RDF. SPARQL anything does not change the SPARQL 1.1 specification but injects new behaviour by overloading the SERVICE operator with a custom IRI-schema. In (Daga et al. 2021), authors introduced the idea of applying façades for semantic lifting, presented a formalisation of the approach in predicate logic, and showed how it is applicable to a variety of formats, while bringing important benefits in terms of usability and extendibility. In (Asprino et al. 2023), it is demonstrated how Façade-X components are enough to be applicable to whatever is generated by a formal grammar (which is the most common tool for describing data formats), and it can in theory be applied to relational databases as well. In this resource paper, we describe the current version of SPARQL Anything and how it was applied to real-world knowledge graph construction pipelines in the context of various projects, including two EU-funded, H2020 projects – SPICE[34] (Daga et al. 2022) and Polifonia[35], and several projects of a US-based IT company active in the healthcare sector. We discuss the value-to-users of our proposition in two ways. First, we report on a survey which evaluates the benefits and opportunities of this approach, compared to alternative solutions, from the user perspective. Second, we provide a field report, presenting first-hand feedback from the industry sector by one of the authors.

The following section illustrates the design principles of façade-based data access (Daga et al. 2021), describing Façade-X, the generic meta-model implemented by our system (Section 2). Section 3 describes the main features of the SPARQL Anything system, the currently supported formats, and additional features that allow the development of RDF construction pipelines from heterogeneous data files. An extensive set of case studies are reported in Section 4. Finally, Section 5 reports on feedback from our community of users. Related work is considered in Section 8. Finally, we conclude the paper in Section 9.

## 2 Façade-based data access

In this Section, we illustrate the idea behind façade-based data access. We rely on the notion of facade as "an object that serves as a front-facing interface masking more complex underlying or structural code"[36]. Applied to our problem, a façade acts as a generic meta-model allowing (a) to inform the development of transformers from an open-ended set of formats, and (b) to generate RDF content in a consistent and predictable way.

The Façade-X meta-model introduced in (Daga et al. 2021) was designed by selecting a small set of primitive data structures: typing, key-value maps, and sequences. Façade-X defines two types of objects: containers and values. Containers can be typed, and one container in the dataset is always of type `root` (the only primitive specified by Façade-X). Values can have any datatype. Containers include a set of unique slots, either labelled as strings or as integer numbers. A slot is filled by another container or by a value. An RDF specification of Façade-X uses the following namespaces and associated preferred prefixes:

```
@prefix rdf:  <http://www.w3.org/1999/02/22-rdf-syntax-ns#> .
@prefix rdfs: <http://www.w3.org/2000/01/rdf-schema#>.
@prefix xsd: <http://www.w3.org/2001/XMLSchema#>.
@prefix fx:   <http://sparql.xyz/facade-x/ns/>. # for fx:root
@prefix xyz:  <http://sparql.xyz/facade-x/data/>. # for the properties
```

The first is used to express the primitive `fx:root`. String slots are RDF properties generated with the `xyz:` namespace, where the local name is supposed to be the string labelling the slot in the data source (for example, a JSON property)[37]. Instead, integer slots (sequences) are represented with instances of `rdf:ContainerMembershipProperty`: `rdf:_1`, `rdf:_2`, ... `rdf:_n`. Façade-X uses containers rather than `rdf:List` since we know that this representation is more efficient to deal with in SPARQL (Daga, Meroño-Peñuela, and Motta 2021). Finally, values are `rdf:Literal`, while containers can be either `IRI`s or blank nodes (the specification does not enforce the use of of either, leaving both options to the Façade-X engineer, including the possibility of switch between the two with a tool option).

---

[34] Social Cohesion, Participation, and Inclusion for Cultural Engagement http://spice-h2020.eu. SPICE is an EU-funded project whose aim is applying a knowledge graph perspective to data exchange and reuse in cultural heritage
[35] Polifonia - A digital harmoniser for Musical Cultural Heritage: https://polifonia-project.eu
[36] https://en.wikipedia.org/wiki/Facade_pattern (accessed, 19/04/2021)
[37] In the case data sources are URI-aware, for example in the case of XML namespaces, the translation engine may reuse the same URIs.

With these set of components, a Façade-X *software* engineer is supposed to design connectors to an open-ended set of resource types, leaving to the *knowledge* engineer (Newell et al. 1982) the freedom of accessing those data sources as-if they were RDF. In what follows, we show how to apply a single RDF abstraction to an extensive set of file formats (including some that are served to the SPARQL users for the very first time).

**CSV**. A CSV file is a resource, identifiable by a URI, which contains text organised as an ordered sequence of rows (newline separated), which in turn contains an ordered sequence of data fields (separated by a delimiter). Rows are ordered; therefore, this case of containment can be represented as an ordered sequence (with container membership properties). What about data values within a row? We observe how CSV data may have an optional "header", where the first line is the list of field names. When this happens, we can use the property component and generate an RDF property reusing the field name, and minting an IRI with the `xyz:` namespace. Otherwise, we can consider the values on each row as another sequence, and fallback to the ordered sequence component.

This is an example from the Tate Gallery open data[38]:

```
id,accession_number,title,…
1035,A00001,"A Figure Bowing",…
…

[ a fx:Root ;
   rdf:_1 [
       xyz:id "1034";
       xyz:accession_number "A00002";
        xyz:title "A Figure Bowing";
… ], … ]
```

**JSON**. The JavaScript Object Notation is specified by ECMA[39]. The syntax defines three types of elements: *objects*, a set of key-value pairs, where keys are supposed to be unique; values, which are either strings, numbers, boolean, or the primitive 'null', and arrays, which specify sequences (containing other arrays, objects, or values). We interpret objects and arrays as containers. We can reuse `rdf:Property` to link objects to values. Arrays can be represented by the ordered sequence component. Values can be expressed as `rdf:Literal`, selecting relevant XSD datatypes from the RDFS specification: `xsd:string`, `xsd:boolean`, `xsd:int`, `xsd:float`[40]. The following example shows a JSON document with metadata of an artist in the Tate Gallery Collection. The JSON file will be represented as follows in RDF (in Turtle syntax):

---

[38] https://github.com/tategallery/collection
[39] https://www.ecma-international.org/publications-and-standards/standards/ecma-404/
[40] Currently, we chose to ignore fields with the 'null' value. However, we may decide to represent it as blank node or to create a primitive entity to express it, for example, similar to rdf:nil.

```
{
  "fc": "Kazimir Malevich",
  "gender": "Male",
  "id": 1561,
  "activePlaces": [ "Ukrayina", "Moskov" ]
}
[ a fx:Root ;
  xyz:fc "Kazimir Malevich",
  xyz:gender "Male"
  xyz:id 1561^^xsd:int,
  xyz:activePlaces [
    rdf:_1 "Ukrayina"; rdf:_2 "Moskov" ]]
```

**HTML and XML**. Both formats can be captured by the Document Object Model (DOM) specification, which we will refer to in the following description[41]. HTML/XML elements (also known as tags) can be definitely considered containers, so we can reuse both the `rdf:Property` Façade-X component for specifying tag attributes, and container membership properties for specifying relations to child elements in the DOM tree. These may include text, which can be expressed as RDF literals of type `xsd:string`. What about element types (tag names)? Façade-X does already provide a solution for *unary* attributes: `rdf:type`. Finally, we can use namespaces declared within the original document to name properties and types, if available, instead of the default `xyz:`. Examples with HTML content will be referred to later in Section 4.

**Textual data** is an interesting case where we can use containment to refer to different elements of the text. The whole content can be included in one single literal, as follows:

```
[] a fx:root ; rdf:_1 "lorem ipsum ..."^^xsd:string .
```

Alternatively, the text can be tokenized (with a user-defined delimiter) and the resulting sequence represented with our façade.

**Binary** content such as images can be also supported, by embedding the content in a single literal of datatype `xsd:binary64encoding`. This solution does not allow to query the actual content, clearly, but still allows to bring in the content and serve it, for example, as linked data. However, binary files such as JPEG images often include annotations, for example, using the common EXIF metadata schema. These annotations can be considered an additional data source, and represented in a separate *metadata* graph with Façade-X.

**YAML** is a lightweight, human-readable data-serialization language. YAML is a "superset" of JSON (any JSON file can be specified in YAML) and, similarly to JSON, data can be organised in lists or associative arrays[42].

**BibTeX** is a text format for computational bibliographies typically used together with the LaTeX system. Each entry consists of the type (e.g. article, inproceedings etc.), a citation key, and key-value pairs for the other characteristics of an entry. Each BibTeX entry can be represented as a typed container that holds a set of key-value pairs.

---

[41] However, it needs to be clarified how our methodology focuses on the elements of the data structure and does not aim at reproducing the DOM API in RDF.
[42] However, differently from JSON, comments and custom data types are allowed. Therefore, in addition to the basic data structures required for capturing JSON files, *rdf:type* is needed for representing custom data types, when available.

A **word processing document** is any text-based document compiled using a word processor software. **Markdown** is a lightweight markup language for writing formatted documents inspired to conventions of web posting. We can interpret a document (compiled with a Word processor or specified in Markdown syntax) as a sequence of blocks (e.g. paragraphs, lists, headings, code blocks). Some blocks (e.g. list items) contain other blocks, whereas others contain inline contents (e.g. links, images etc.). A document can be represented as a list of typed containers. Where the type denotes the kind of block (e.g. heading, paragraph, emphasised text, link, image etc.); *lists* are needed for specifying the sequence of the blocks. Additional attributes such as the depth of the header or the type of list (bullets, numbers, etc...) can be also supported, relying on the key-value structure. The following shows an example of Markdown:

```
# SPARQL Anything
SPARQL Anything is a system for Semantic Web re-engineering that
allows users to ... query anything with SPARQL.
…
[] a fx:root ; a xyz:Document ;
 rdf:_1 [ a xyz:Heading ;
  rdf:_1 "SPARQL Anything"^^xsd:string ;
  xyz:level "1"^^xsd:int ] ;
 rdf:_2 [ a xyz:Paragraph ;
   rdf:_1 "SPARQL Anything is ..."^^xsd:string  ] .
…
```

**Directory structures** can be interpreted as collections of folders (containers) or file names (values). This allows to develop façade-based data access to explore the content of a local directory. The same concept applies to archives (e.g. zip files).

As discussed, `Facade-X` is the same for all the surveyed formats. façade-based data access acts as a virtual endpoint that can be queried exactly as a remote SPARQL endpoint, through a SERVICE call to the special IRI-schema `x-sparql-anything:`. The related URI-schema supports an open-ended set of parameters specified by the facade implementations available. A minimal example only includes the resource locator, and guesses the data source type from the file extension. Options are embedded as key-value pairs, separated by comma. These can incorporate a set of parameters, to allow the user to configure the system (for example, to indicate that the system should consider the first line of a CSV as headers):

```
x-sparql-anything:media-type=application/json; charset=UTF-8,loca
tion=...
```

Figure 1 shows an example query. The query (taken from the Tate Gallery Showcase [Daga 2022]) selects artwork metadata from a CSV file and collects additional data from a related JSON file from the local directory. First, it iterates over a CSV with artworks' metadata and, for each one of them, constructs the path to the local JSON file. Then, the JSON file is queried for artwork subjects. The query solution is finally projected into a CONSTRUCT clause[43].

---

[43] This example is reproducible, along with other example queries on the same data source, at https://github.com/sparql-anything/showcase-tate

## 3  SPARQL Anything: system overview

In what follows, we describe the process of executing façade-based data access with SPARQL Anything. After that, we summarise the main features of the system.

SPARQL Anything extends the Apache Jena framework[44] with a special query executor capable of handling façade-based data access. The system behaves essentially as a standard SPARQL 1.1 query engine, receiving as input a query and returning either a SPARQL Result Set (for SELECT/ASK queries) or an RDF stream (for CONSTRUCT/DESCRIBE types of queries). Figure 2 describes the general architecture of the system.

```
PREFIX fx: <http://sparql.xyz/facade-x/ns/>
PREFIX xyz: <http://sparql.xyz/facade-x/data/>
PREFIX rdf: <http://www.w3.org/1999/02/22-rdf-syntax-ns#>
PREFIX tate: <http://sparql.xyz/example/tate/>
PREFIX tsub: <http://sparql.xyz/example/tate/subject/>
PREFIX schema: <http://schema.org/>
PREFIX dct: <http://purl.org/dc/terms/>
PREFIX rdfs: <http://www.w3.org/2000/01/rdf-schema#>

CONSTRUCT {
  ?artwork a schema:CreativeWork ;
    dct:subject ?subject ;
    schema:thumbnailUrl ?thumbnail ;
    dct:title ?title .
  ?subject rdfs:label ?subjectName .
} WHERE {
  # List of artworks
  SERVICE <x-sparql-anything:csv.headers=true,location=./collection/ar
twork_data.csv> {
    []   xyz:id ?id ; xyz:accession_number ?accId ;
         xyz:title ?title; xyz:thumbnailUrl ?thumbnail .
  }
  BIND (IRI(CONCAT(STR(tate:), "artwork-", ?id )) AS ?artwork) .
  BIND (IRI(CONCAT(STR(tate:), "artist-", ?artistId )) AS ?artist) .
  # Build file path to JSON file for each record
  BIND ( IF ( STRSTARTS( ?accId, "AR" ) ,
    LCASE(CONCAT( "ar/", SUBSTR( ?accId ,3 ,3), "/", ?accId, "-", ?id
, ".json" )),
    LCASE(CONCAT( SUBSTR( ?accId ,1 , 1), "/", SUBSTR( ?accId ,2 , 3),
"/" , ?accId, "-", ?id , ".json" ))
  ) AS ?filepath ) .
  # JSON file with subjects
  BIND (fx:entity("x-sparql-anything:location=./collection/artworks/",
?filepath ) AS ?artworkMetadata ) .
   SERVICE ?artworkMetadata {
         # Query topics in JSON files
         [] xyz:children [ fx:anySlot [ xyz:id ?subjectId ] ].
   } .
  BIND (fx:entity(tsub:, ?subjectId )) AS ?subject) .
}
```

*Figure 1. Example of SPARQL Anything query from the Tate Gallery Collection showcase (Daga 2022). The query selects artworks' metadata from a CSV, builds the path to a related JSON file containing additional annotations (subjects). These JSON files are queried in another façade-based data access operation, where subjects are collected. The variables are projected into a Knowledge Graph design in the CONSTRUCT clause.*

---

[44] https://jena.apache.org/index.html

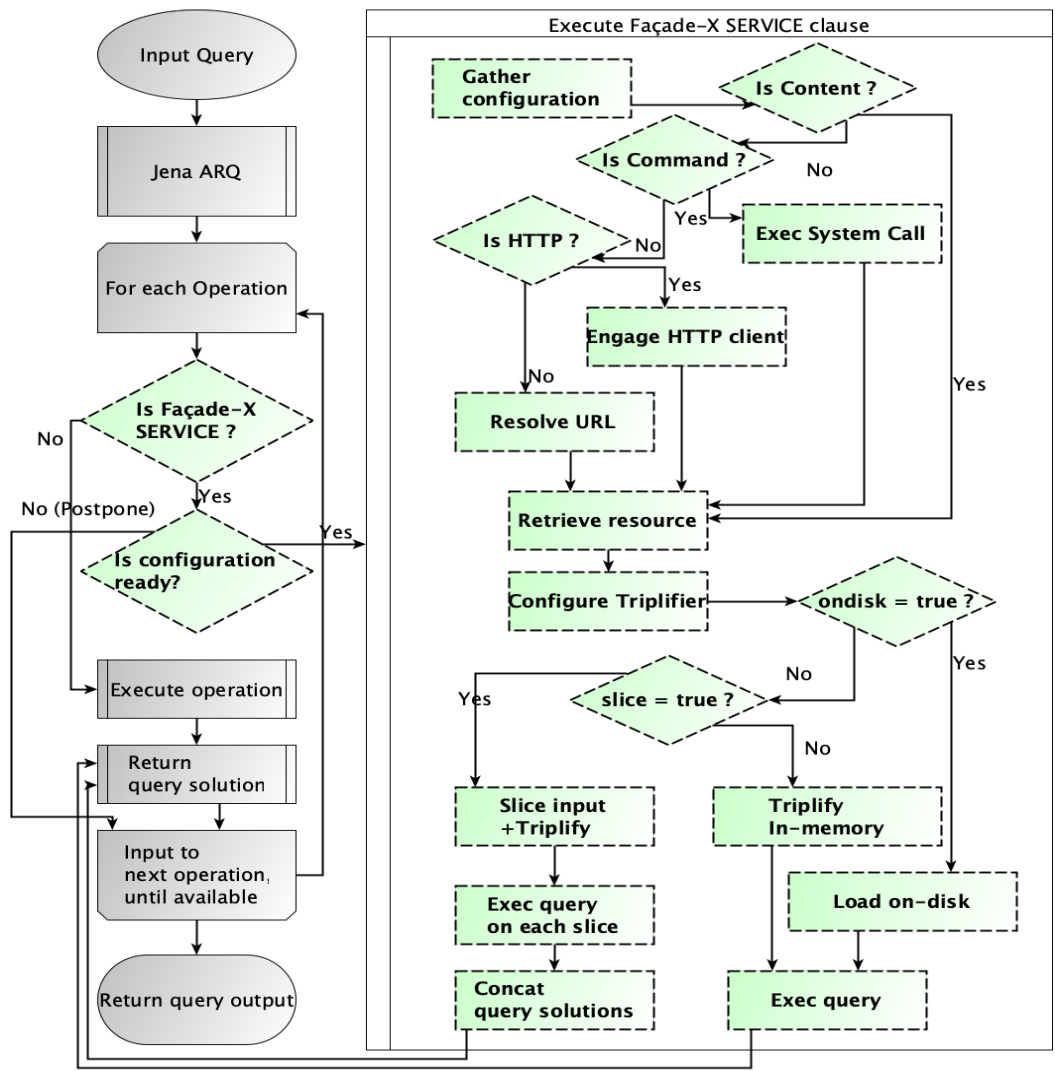

*Figure 2. SPARQL Anything: system architecture.*

The process starts with an input query, which is handled by the ARQ engine of Apache Jena. The query is parsed into an abstract algebra, and operations are executed according to their internal dependencies, where the output of one operation is served as input to the dependent one. Our system intercepts attempts to execute any SERVICE pointing to a `x-sparql-anything:` IRI. However, configuration can be expressed to SPARQL Anything either via the IRI schema (as in the previous example query) or by using triples having as subject the special entity `fx:properties`, as in the following examples:

```
fx:properties fx:location "./my-file.csv" .
[or]
fx:properties fx:location "http://my-web-api" ; fx:media-type "application/json" ; fx:http.query.param.api-key "my-api-key" .
[or]
fx:properties fx:command "echo first,second,third" ; fx:media-type "text/csv"
[or]
fx:properties fx:content "first,second,third" ; fx:media-type "text/csv"
[or]
fx:properties fx:content ?data ; fx:media-type "text/csv"
```

In the last line, the input data is supposed to come from a previous operation. Therefore, if there are configuration variables that are not been evaluated, the execution is postponed. When all the input parameters are ready, our process starts. The system first gathers all configuration options. We refer the reader to the official documentation on the web for a complete list of configuration options, including format-specific ones ("SPARQL Anything software documentation" 2022). Next, the system identifies the input source. Currently, SPARQL Anything supports input from within the SPARQL query – content, or by defining a command to be executed on the host machine (process STDOUT is streamed to the triplifier), or by specifying a resource URL – option *location*. In the case the location is an HTTP URL, the resource is resolved via a full fledged HTTP client. HTTP client options include setting the HTTP method (GET< POST, PUT,...), passing authentication options and query parameters, and setting HTTP request headers such as `Accept: application/json; charset=utf-8`. Through this component, SPARQL Anything is capable of supporting querying complex Web APIs. The output of the request can be interpreted as any of the content types supported. Finally, any other URLs are resolved via the underlying Java IO URL connection library. Independently from the method to obtain a resource, the input is passed to a triplifier. In this phase, the system uses the option *media-type* or tries to guess the format from the file extension, if available. The re-engineering of the input is performed by the triplifier according to the available configuration options. By default, the process triplifies all the content, materialising a view *in-memory*, then executes the query (the inner part of the SERVICE clause) and returns a *query solution*, like any other SPARQL operations. When the input is too large to be loaded all in-memory, the user has two possibilities. The *on-disk* option instructs the system to save the façade-based RDF into a temporary local triple store database (Jena TDB2), and then execute the query on the database. In the alternative, the *slice* option can be invoked, and the content is partitioned and the triplification and query execution performed on each one of the parts separately (currently supported only for CSV, JSON, and XML). The output is streamed together so that external operations can continue in the same way as with an execution with a complete view. In all cases, the triplification process only preserves the part of the Façade-X RDF view that is mentioned in the input query – *triple-filtering*. Specifically, if no quad pattern in the SERVICE clause matches a given generated RDF statement, this is not included in the materialised view (this feature, enabled by default, can be disabled via the option *strategy*, see the documentation ("SPARQL Anything software documentation" 2022).

We illustrate the main features of SPARQL Anything, referring to the official documentation for further details ("SPARQL Anything software documentation" 2022). Users can customise the behaviour of façade-based data access according to a set of options. In addition to the ones already mentioned, options include preparing string values (`trim-strings`), generate IRIs instead of BNodes (`blank-nodes=false`), specifying the input `charset`, or a custom `namespace` instead of the default `xyz:`. With the option `null-string`, users can request to ignore triples having as value the given string (for example, an empty string or the string `"N/A"`). Additional metadata can be extracted (e.g. from EXIF annotations in image files) via the `metadata=true` option.

The system allows to query the following file formats in plain SPARQL 1.1: XML, JSON, CSV, HTML, Excel, Text, Binary, EXIF, File System, Zip/Tar, Markdown, YAML, Bibtex, DOCx. Users can customise the behaviour of the triplifiers with format-specific options. For example, JSON can be filtered passing a JsonPath expression (`json.path`), while XML with XPath (`xml.path`), HTML content with a CSS selector (`html.selector`), and plain text via a regular expression (`txt.regex`) or a string delimiter (`txt.split`). The CSV triplifier can be applied to any char-separated format (e.g. TSV) with the option `csv.delimiter`. HTML pages can be loaded with a virtual browser before querying, allowing to parse content produced via JavaScript (`html.browser`). SPARQL Anything includes a full fledged HTTP client to query Web APIs (options include `http.header.*`, `http.auth.*`, `http.query.*`, and others). SPARQL Anything exposes an extensive set of functions in addition to the ones already provided by Apache Jena, for example, the magic property `fx:anySlot` to match the value of any container membership property. These include shorthand functions for building RDF nodes – `fx:entity`, `fx:literal` – and querying and manipulating container membership properties (`fx:before`, `fx:after`, `fx:prev`, `fx:next`), string manipulations and hashes, and others (Wood et al. 1998).

SPARQL Anything can be used as a Command Line Interface or via as a SPARQL Endpoint, featuring Apache Jena Fuseki. The CLI supports additional features that allow to combine the output of a SPARQL Anything query as the input of another one, designing rich data flows. Output formats (`-f`) can be JSON, XML, CSV, TEXT, TTL, NT, NQ, and result can be saved to a file (`-o`). For example, query in Figure 1 can be executed as follows:

```
fx -q queries/arts-and-subjects.sparql -f TTL -o arts-and-subjects.ttl
```

The system supports parametrized queries using the BASIL syntax convention (Daga, Panziera, and Pedrinaci 2015). Users can specify a SPARQL Result Set file to provide variable parameter values (option `-i`) or specify values inline (`-v`). When present, the query is pre-processed by replacing parameters with values from the input file (or values), and repeated for each of the provided bindings. Parameter values can be used in the output file name (`-p`). In addition, it is possible to reuse content from a previously performed transformation and execute the query against an existing (set of) RDF files (option `-l` or `--load`). The option requires the path to one RDF file or a folder including a set of files to be loaded. When present, the data is loaded in memory and the query executed against it.

## 4  Showcase

In this Section we provide references to a set of real-world use cases implemented with SPARQL Anything.

*The Tate Gallery Collection open data*

This showcase (Daga, 2022) provides examples of using SPARQL Anything to query open data from the Tate Gallery collection (Daga, 2022). The repository shows basic features such as the use of the option `csv.headers=true`. In addition, it showcases more advanced features demonstrating how to (a) query local CSV and JSON files together to build a knowledge graph of artists and artworks metadata using Schema.org and (b) build a SKOS taxonomy from artworks' topics distributed in thousands of JSON files. Showcased features include incorporating binary data in the RDF graph and dynamically generating the *location* option from previous façade-based data access operations, among others.

*The Irish Museum of Modern Art Website*

Data can be anywhere! In this showcase, implemented during the SPICE project (Daga et al. 2022), we developed a knowledge graph of artists and artworks scraping the website at http://imma.ie. The code shows how to query an HTML web page using a custom CSS selector (option `html.selector`), and the use of BASIL (Daga, Panziera, and Pedrinaci 2015) variables in parametrised queries. In addition, the use case developed demonstrates how CLI commands can be used in sequence to build complex knowledge graph construction pipelines. Through these features, we extract artists names and pages, from which we scrape additional metadata, including the list of artworks' pages'. These are visited next to complete the KG with artworks' information.

*Generating a Knowledge Graph from The Proposition Bank (PropBank)*

This showcase features the popular PropBank corpus of *linguistic frames*[45]. The input for the process is a release of the PropBank dataset, typically released as a single zip file containing a folder which stores all the XML files, one for each *frame*. The query shows how to chain multiple façade-based data access operations, exploiting key features of SPARQL Anything such as archive and directories querying (to get all XML files from archive), iterate on the solution of one operation to feed it into another (query each one of the XML) and project all transformations into a CONSTRUCT graph template.

*Scraping Webpages with SPARQL*

This online guide[46], explores advanced usage of the HTML triplifier, showing features such as the headless browser. In addition, it demonstrates the use of list functions such as `fx:before`, `fx:next`, and others.

*Querying YAML metadata embedded in GitHub Markdown files*

Projects on GitHub typically include a `README.md` file. The Polifonia project publishes an *ecosystem* of tools and data for the computational treatment of musical cultural heritage. Such collection of material is spread over a number of GitHub repositories exposing a collection of *components*, each one described in a documentation Markdown file, annotated according to an *annotation schema*. This showcase demonstrates how to query Markdown and YAML annotations contained within by chaining multiple façade-based data access operations with a single SPARQL Anything query, exploiting the *content* option. The query[47] traverses a local file system (where all the relevant repositories are included) in search of .md files, extracts the YAML front matter, and transforms the annotations according to Façade-X . Relevant RDF is projected into a KG of component types.

*Musical scores feature extraction (MusicXML)*

Musical scores are an excellent example of a complex data object. In (Ratta and Daga, 2022), the authors explore the application of SPARQL Anything to extrapolate musical features such as (a) extracting melodic information, (b) extracting N-grams of musical information, (c) supporting the analysis of those N-grams and (d) populate a musical note ontology (code available on GitHub[48]).

## 5 Engagement with target users

A survey was conducted in order to engage with Semantic Web practitioners and SPARQL developers and users and also gain some initial insights into their data transformation requirements and perception and use of existing tools. There were 27 completed responses to the survey. A fuller account of the survey can be found in (Daga et al. 2021). Participants covered a diverse range of expertise. 37% rated their expertise in transforming data to RDF as high or very high, and 36% as low or none. 51.8% were frequent or very frequent users of SPARQL 1.1. 33.3% rated their expertise in SPARQL 1.1. as high or very high, and 37% as low. The software most commonly used for data transformation was custom code written for the specific task, followed by RML and SPARQL Generate. Participants transformed datasets of different sizes, from less than 10MB (18.5% of participants) through to over 1GB (also 18.5% of participants). 33% transformed the data into fewer than 1 million triples. 22.2% generated over 100 million triples.

---

[45] https://propbank.github.io
[46] https://github.com/justin2004/weblog/tree/master/scraping_with_sparql
[47] https://github.com/SPARQL-Anything/showcase-polifonia-ecosystem/blob/master/queries/components-to-rdf.sparql
[48] https://github.com/SPARQL-Anything/showcase-musicxml

A set of questions in the survey asked participants to rate the importance of various usability characteristics of systems for transforming data into RDF. 51.8% considered it very important or essential that the system should minimise the languages and syntaxes needed. 70.3% considered it very important or essential that the system be easy to learn. 55.5% considered it essential or very important that the system can support new types of data sources without changes to the mapping language.

A final set of questions asked participants to rate the usability of three notations for transforming non-RDF data into RDF: RML, SPARQL Generate and SPARQL Anything. 29.6% rated data transformations specified in SPARQL Anything code as very easy to understand, 63% as easy to understand. 4.8% rated the SPARQL Generate code as very easy, 40.7% as easy. 7.4% rated the RML code as very easy, 22.2% as easy.

In summary, this small survey of a sample of target users covering a range of expertise and experience suggests a potentially receptive audience for an easy-to-use method for transforming a range of data sources without the need to learn additional languages and syntaxes.

## 6  The Open-Source project

SPARQL Anything is produced by a community of contributors as open source software distributed under the commercial-friendly Apache Licence 2.0. The project is managed on GitHub at this address: https://github.com/SPARQL-Anything, and can be cited via its related entry in Zenodo ("SPARQL Anything software" 2022), following good practices of Open Science and FAIR data management policies. The official documentation is published via Readthedocs.io[49]. The development activity started in November 2020 and has continued since then with a steady increment of contributions from people external to the original team, including practitioners from the industry. The GitHub project has currently 12 watchers, 5 forks, and 132 stars. In one month (March 2023, time of this submission), the project had 72 unique visitors and was cloned by 25 unique users.

## 7  The industry perspective: a field report

In this Section we report on the experience of a software engineer who joined the SPARQL Anything open source project recently, working within the context of an US-based IT company active in the healthcare sector. The US company team included, apart from software engineers, two ontologists and a data scientist, all collaborating for constructing RDF KGs for about a year and a half. It took the team several months to find a method for constructing RDF that worked well for them. Initial experiments involved R2RML/RML based tools, including Mapeathor[50] (Iglesias-Molina et al. 2020), RML Mapper[51], and Ontop[52] (Calvanese et al. 2017).

At first, the ontologists did not implement their own mappings. Instead, they would annotate sample data with the triples they would like to produce and then one of the software engineers would use one of the RML based tools to implement the mappings. However, the ontologists were unable to revise the mappings and any subsequent evolution required intervention by the software engineers. Indeed, SPARQL and Bash were the only common languages everyone on the team used. The team was changing tools on each project, in search for a better solution.

---

[49] https://sparql-anything.readthedocs.io/en/latest/
[50] https://github.com/oeg-upm/mapeathor
[51] https://github.com/RMLio/rmlmapper-java
[52] https://github.com/ontop/ontop

The discovery of SPARQL Anything had a huge impact on the workflow, allowing the team to transform more sources and doing it more quickly. Now the ontologists and data scientist implement their own mappings and everyone participates in the maintenance of the mappings.

One episode illustrates well the impact that the tool had on the team. A few months ago, the company needed graph of data from a paying service to which the company had access. Unfortunately, the data was only available via a web browser as HTML. This brought the opportunity of contributing to the SPARQL Anything open-source project by expanding the capabilities of the HTML triplifier with a headless browser option, illustrated in a blog post (see Section 4), written with the ontologists in mind, demonstrating how to scrape a webpage (with content produced by javascript) using SPARQL Anything[53]. As a result, the ontologists read the blog post and created a SPARQL construct query, without any assistance from the software engineers, to construct a KG with the content of the webpage of interest.

Recently the team were involved in a short workshop where one of the goals was to produce a graph of a product catalog[54]. Because the team was producing triples so quickly the workshop was mainly spent working with a subject matter expert on the data in order to *carve nature at its joints.* The team was happy to adapt their workflow so that ontologists could use SPARQL Anything in complex data integration pipelines, sometimes involving millions of data points. In one case, healthcare data is integrated from a relational database that gets exported (as CSV) weekly. The ontologists wrote one or more construct queries for each table. To overcome the in-memory limits of the tool, a script split each CSV file into files of a few thousand rows. SPARQL Anything was run over each file (with a configurable number of parallel processes) to produce quads. As an example, with 5 parallel processes one table with about 10 million rows took 4-5 hours to complete and produced 82 million quads. The team used this technique before SPARQL Anything had the *slice* option. As that option gets more optimised, the team looks forward to having SPARQL Anything doing the slicing rather than doing it as a pre-processing stage.

## 8 Related work

Motivation for our work resides in research on end-user development and human interaction with data. End-user development is defined by (Lieberman et al. 2006) as "*methods, techniques, and tools that allow users of software systems, who are acting as non-professional software developers, at some point to create, modify or extend a software artefact*". Many end-user development tasks are concerned with the use of software to manipulate data. Recent works encompass sending, receiving and manipulating data from web APIs, IoT devices and robots (Paternò and Santoro 2019). Unlike professional software development, end-user development involves the construction of software for personal rather that public use (Ko et al. 2011) in order to carry out professional activities. Many SPARQL users fall into the category of end-user developer. In a survey of SPARQL users, (Warren and Mulholland 2018) found that although 58% came from the computer science and IT domain, other SPARQL users came from non-IT areas, including social sciences and the humanities, business and economics, and biomedical, engineering or physical sciences. In addition, findings in this area (Panko and Aurigemma 2010) suggest that the data with which users work is more often primarily list-based and/or hierarchical rather than tabular. For example, (Chang and Myers 2016) proposes an extension to spreadsheets to explicitly support hierarchical data and (Hall 2019) proposes an alternative formulation to spreadsheets in which data is represented as *list-of-lists*, rather than tables. Our proposal goes in this direction and accounts for recent findings in end-user development research.

---

[53] https://github.com/justin2004/weblog/tree/master/scraping_with_sparql
[54] Technical aspects were reported in https://github.com/justin2004/weblog/tree/master/SPARQL_value_functions

We survey approaches to extend SPARQL. A standard method for extending SPARQL is by providing custom functions to be used in FILTER or BIND operators[55]. Query processing engines can extend SPARQL by using so-called magic properties. This approach defines custom predicates to be used for instructing specific behaviour at query execution[56]. SPARQL Generate (Lefrançois, Zimmermann, and Bakerally 2017) introduces a novel approach for performing data transformation from heterogeneous sources into RDF by extending the SPARQL syntax with a new GENERATE operator (Lefrançois, Zimmermann, and Bakerally 2017). The method introduces two more operators, SOURCE and ITERATOR. Custom functions perform ad-hoc operations on the supported formats, for example, relying on XPath or JSONPath. However, there are also approaches to extend SPARQL without changes to the standard syntax. For example, BASIL (Daga et al. 2015) allows to define parametric queries by enforcing a convention in SPARQL variable names. SPARQL Anything reuses BASIL variables to support parametric queries and file names. SPARQL Micro-service (Michel et al. 2019) provides a framework that, on the basis of API mapping specification, wraps web APIs in SPARQL endpoints and uses JSON-LD profile to translate the JSON responses of the API into RDF. In this paper, we follow a similar, minimalist approach and extend SPARQL by overriding the behaviour of the SERVICE operator.

Several tools are available for automatically transforming data sources of several formats into RDF (Any23[57], JSON2RDF[58], CSV2RDF[59] to name a few). While these tools have a similar goal (i.e. enabling the user to access the content of a data source as if it was in RDF), the (meta)model used for generating the RDF data highly depends on the input format thus limiting the homogeneity of data generated from heterogeneous data formats. However, none of these approaches are based on a common abstraction to heterogeneous formats.

Mapping languages for transforming heterogeneous files into RDF are represented by RML (Dimou et al. 2014), also specialised to support data cleaning operations (Slepicka et al. 2015) and specific forms of data, for example, relational (Rodriguez-Muro and Rezk 2015) or geospatial data (Kyzirakos et al. 2014). This family of solutions are based on a set of declarative rules that Semantic Web practitioners are expected to develop by analysing the input data sources. The language incorporates format-specific query languages (e.g. XPath) and require the practitioner to have deep knowledge not only of the input data model but also of standard methods used for its processing. A recent approach (Beret, Papadakis, and Koubarakis 2020) applies OBDA to SPARQL any type of Web resource, with a sophisticated set of mappings supporting an intermediate query service. Authors of a recent alternative, based on SHeX (García-González et al. 2020), stress the importance of making mappings usable by end users. Indeed, recent work acknowledges how these languages are built with machine-processability in mind (Heyvaert et al. 2018) and how defining or even understanding the rules is not trivial to users. SPARQL Anything goes beyond current approaches and aims at equipping SPARQL users with the simplest possible assumption on how to deal with heterogeneous resources.

---

[55] ARQ provides a library of custom functions for supporting aggregates such as computing a standard deviation of a collection of values. ARQ functions: https://jena.apache.org/documentation/query/extension.html (accessed 15/12/2020).
[56] For example, this allows the specification of complex fulltext searches over literal values. Query processors can delegate execution to a fulltext engine (e.g. Lucene) and return a collection of query solutions as triple patterns
[57] http://any23.apache.org/
[58] https://github.com/AtomGraph/JSON2RDF
[59] http://clarkparsia.github.io/csv2rdf/

We have already discussed how the SPARQL Anything approach is doing better than alternative solutions, from the SPARQL user standpoint. However, this comes at the cost of leaving the internal machinery to deal with the difficulty of developing strategies for executing SPARQL queries on different formats. For example, our survey highlighted the importance of being able to cope with very large data sources, partly addressed by the on-disk and slice options. Future work includes research on query-rewriting approaches to stream the data, similarly to the internal machinery of SPARQL Generate. Actually, one way of doing it would be rewriting the queries from the Facade-X structure to SPARQL Generate expressions, and run them as a back-end component.

Currently, our system supports files, Web APIs, and content streamed from a separate process. Future work includes support the connection to relational databases, for example, relying on engines implementing the W3C Direct Mapping recommendation (Prud'hommeaux et al. 2012) for relational databases. To do that, we aim at reusing recent research in OBDA (e.g. [Sequeda and Miranker 2017]) to develop optimised query-rewriting strategies from Facade-X to the underlying relational model, on demand, without asking users to engage with the mappings. We plan to implement different methods, including experimenting with alternative back-end engines.

## 9 Conclusions

We introduced SPARQL Anything, a reusable research software supporting façade-based data access over heterogeneous data sources, for the benefits of knowledge engineers. The Façade-X approach, introduced in (Daga et al. 2021) has received the attention of the community, confirmed by the user survey (Section 5), and it is unique in the landscape of solutions for RDF re-engineering. SPARQL Anything has a low complexity, compared to existing solutions: the framework allows others to save significant coding effort. In addition, the design methodology allows it to be extended to support an ever-ending set of formats. The performance of the first version of the system was evaluated and discussed in (Daga et al. 2021). Future work is going to focus on optimisation strategies and improving the internal machinery. However, the current version of the system is ready to tackle real-world problems, such as the ones encountered in our field report. In addition, engaging with semantic web practitioners highlighted the *value-to-users* of essential and unique aspects of our tool. With SPARQL Anything, Semantic Web practitioners are relieved of the problem of content re-engineering and can finally focus on generating high-quality semantic data in plain SPARQL.

## 10 Acknowledgements

This work was partially supported by the EU's Horizon Europe research and innovation programme within the SPICE project (Grant Agreement N. 870811) and the Polifonia project (grant agreement N. 101004746).